\documentclass[twocolumn,showpacs,pra,amsmath]{revtex4}
\usepackage{epsfig,graphicx,bm}

\def\be{\begin{equation}}
\def\ee{\end{equation}}
\def\bea{\begin{eqnarray}}
\def\eea{\end{eqnarray}}
\def\ea#1{\label{#1}\end{eqnarray}}
\def\bes#1{\begin{subequations}\label{#1}}
\def\ese{\end{subequations}}

\newcommand{\abs}[1]{\left| #1 \right|} 

\begin{document}
\title{Simple quantum error detection and correction for superconducting
qubits}
\author{Kyle Keane and Alexander N.\ Korotkov}
 \affiliation{Department of
Electrical Engineering and Department of Physics \& Astronomy,
University of California, Riverside, California 92521}
\date{\today}

\begin{abstract}
We analyze simple quantum error detection and quantum error
correction protocols relevant to current experiments with
superconducting qubits. We show that for qubits with energy
relaxation the repetitive $N$-qubit codes cannot be used for quantum
error correction, but can be used for quantum error detection. In
the latter case it is sufficient to use only two qubits for the
encoding. In the analysis we demonstrate a useful technique of
unraveling the qubit energy relaxation into ``relaxation'' and ``no
relaxation'' scenarios.
 Also, we propose and numerically analyze several two-qubit algorithms
for quantum error detection/correction, which can be readily
realized at the present-day level of the phase qubit technology.
    \end{abstract}
\pacs{03.67.Pp, 03.67.Lx, 85.25.Cp}


  \maketitle

\section{Introduction}

    Quantum error correction \cite{QEC-theory} (QEC) is an unavoidable
procedure in a practical quantum computer \cite{N-C,Preskill-98}.
The standard QEC \cite{QEC-theory,N-C} includes encoding a logical
qubit in several physical qubits, measuring the error syndrome using
ancillary qubits, and then applying a correction operation, which
depends on the measurement result. (A promising variation of this
idea are the so-called surface codes \cite{surface-codes}.)
Unfortunately, QEC is very difficult experimentally
\cite{Cory-98,Knill-01,Cory-05,Moussa-11,Chiaverini-04,Blatt-11,
Pittman-05,Aoki-09,Lassen-10,Yao-12,Reed-12}, and some
simplifications are often used. Let us mention three of them, all of
which have been introduced in Ref.\ \cite{Cory-98}. First, instead
of using additional qubits for the error syndrome, in a ``compact''
scheme the same physical qubits can be  used for the encoding and
error syndrome measurement; this is done by decoding the encoded
state after a possible error occurs. Second, since a single-shot
measurement of a qubit state is often difficult, the standard QEC
can be replaced by measurement-free QEC, in which the measurement
and correction are substituted by a quantum controlled operation
(e.g.\ the Toffoli gate). Third, a favorable type of error (against
which the code protects) is often simulated by applying a certain
unitary rotation, with the rotation angle corresponding to the error
strength.

Measurement-free QEC experiments in nuclear magnetic resonance (NMR)
systems \cite{Cory-98,Knill-01,Cory-05,Moussa-11} have been
performed for over a decade, but only with ensembles of quantum
systems \cite{Caves-02}. Using trapped ions, a three-qubit QEC
experiment with actual measurement was realized
\cite{Chiaverini-04}, and recently a measurement-free QEC procedure
with several error correction cycles was demonstrated
\cite{Blatt-11}. In linear optics systems, the QEC experiments
include two-qubit protection against ``accidental'' measurement
\cite{Pittman-05}, continuous-variable adaptation of the 9-qubit
Shor's code \cite{Aoki-09}, continuous-variable erasure-correcting
code \cite{Lassen-10}, and eight-photon topological error correction
\cite{Yao-12}. A three-qubit measurement-free QEC protocol has been
recently demonstrated with superconducting ``transmon'' qubits
\cite{Reed-12}.

    With the rapid progress in experiments with superconducting
qubits \cite{SCQ-review,QIP-issue}, QEC with actual measurements is
becoming feasible in these systems in the reasonably near future.
The subject of this paper is the analysis of several simple quantum
error correction/detection protocols relevant to future experiments
with superconducting qubits, mainly superconducting phase qubits
\cite{Martinis-09}. (Some results of this paper have been reported
earlier \cite{Keane-QEDC}.)

In the past, pure dephasing was by far the dominant source of
decoherence in superconducting qubits, and QEC protecting against
pure dephasing would be most important. An example of such a
procedure was considered theoretically in Ref.\ \cite{Shumeiko}. The
idea was to use the standard 3-qubit repetitive code, which protects
from bit flips (i.e.\ $X$-rotations). By using additional Hadamard
gates for each physical qubit, the $X$-rotations are converted into
$Z$-rotations, and therefore the same code can be used to protect
against pure dephasing.

    In recent years, pure dephasing in superconducting qubits
was significantly reduced by various technological advances
\cite{SCQ-review,QIP-issue}, and now energy relaxation is becoming
most important. In particular, when quantum information is stored in
a superconducting resonator \cite{Johnson-10,Matteo-11}, pure
dephasing is negligible in comparison with energy relaxation. This
is why in the first part of this paper (Sec.\ II) we focus on the
operation of repetitive $N$-qubit quantum codes in the presence of
energy relaxation. Repetitive codes are chosen because of their
relative simplicity in the encoding and decoding (unfortunately, the
standard 5-qubit or 7-qubit stabilizer codes \cite{QEC-theory, N-C,
Laflamme-96} are not feasible for superconducting qubits in the near
future). To reduce the number of qubits in a procedure we use the
standard compact scheme \cite{Cory-98,Shumeiko}, in which the
ancilla qubits used for encoding are also used for the error
syndrome measurement. We assume that the energy relaxation happens
at zero temperature, which is essentially the case for
superconducting phase qubits, since the typical qubit frequency is
$\sim$6 GHz, and therefore the energy $\hbar\omega \simeq 0.3$ K is
much larger than the experimental temperature of $\sim$50 mK.

 Even though energy relaxation may look similar to a bit-flip, it
actually can be thought of as a combination of two quantum errors:
bit-flip and bit-phase-flip (which correspond to $X$-rotation and
$Y$-rotation). This is the reason why, as we show later explicitly,
repetitive codes do not work for QEC against energy relaxation.
However, these codes can be efficiently used for quantum error
detection (QED). In QED we detect that an error happened but cannot
restore the undamaged quantum state (in particular, the QED idea was
implemented in Ref.\ \cite{Leung-99} for phase errors in
liquid-state NMR and has been recently investigated in Ref.\
\cite{Lassen-10} for detecting photon erasures). Even though QED is
of much more limited use than QEC, it is still an interesting
procedure, and experimentally can be considered as a first step
towards full QEC. We show that for QED against energy relaxation it
is sufficient to use 2-qubit encoding and that there is not much
benefit to use more qubits, unless a somewhat more sophisticated
procedure is used.

    Our analysis in Sec.\ II is based on unraveling the qubit energy
relaxation into the ``relaxation'' and ``no relaxation'' scenarios.
This unraveling is quite different (and more difficult) than, for
example, unraveling of pure dephasing into the ``phase flip'' and
``no phase flip'' scenarios. The main reason for the difference is
that the unraveled states for the energy relaxation are related to
the initial state in a non-unitary way.

    In Sec.\ III we focus on simple two-qubit QEC/QED protocols,
somewhat similar to those in Ref.\ \cite{Leung-99}, which can be
readily implemented using the present-day technology of phase qubits
\cite{Matteo-11}. Realistic experimental parameters are used in the
numerical simulation of these protocols.
    In the first protocol and its variations, we assume that as
in most of the previous experiments
\cite{Cory-98,Knill-01,Cory-05,Moussa-11,Chiaverini-04,Blatt-11,
Pittman-05,Aoki-09,Lassen-10,Yao-12,Reed-12} the errors are
intentionally induced by particular operations pretended to be
unknown. The algorithms can mainly be used for QED; however, when
the type of particular error is known (which is the case for
intentional errors in an experiment), the algorithms can also be
used for QEC. We also analyze numerically the operation of a
protocol, in which the errors during a storage period are due to
actual energy relaxation of two qubits (assuming storage in
resonators of a RezQu-architecture device \cite{Matteo-11,RezQu}
with phase qubits). This protocol can only be used in the QED mode.
The main result of the simulations is that the analyzed protocols
can be realized at the present-day level of phase qubit technology.
    Section IV is a conclusion. Some mathematical details of the analysis
are discussed in the Appendix.

\section{Repetitive coding for energy relaxation}

    In this section we analyze the operation of repetitive $N$-qubit
encoding in the presence of (Markovian) zero-temperature energy
relaxation. The procedure is shown in Fig.\ 1. The goal is to
preserve an arbitrary initial state
    \be
|\psi_{\rm in}\rangle= \alpha |0\rangle +\beta |1\rangle
    \ee
of the main (upper) qubit, where $|0\rangle$ is the ground state and
$|1\rangle$ is the excited state. In this paper we consider only
preservation of the initial state (``memory'' operation), so in
discussing the fidelity of a procedure we always imply comparison
with the ideal memory operation.

The encoding in Fig.\ 1 is performed with $N-1$ controlled-NOT
(CNOT) gates, acting on $N-1$ ancilla qubits, which all start in the
state $|0\rangle$. This produces the $N$-qubit wavefunction $\alpha
|0^N\rangle +\beta |1^N\rangle$, where the notation $|x^N\rangle$
represents the product-state of $N$ qubits, all being in the state
$x$. After the encoding, all qubits are subjected to decoherence due
to zero-temperature energy relaxation with relaxation time
$T_{1}^{(i)}$ for the $i$th qubit, $i=1,2,\dots N$. We will mostly
consider the case when the decoherence is the same for all qubits,
$T_{1}^{(i)}=T_1$. After the decoherence during time $t$, the logic
state is decoded by using $N-1$ CNOT gates in the same way as was
done for the encoding, and all $N-1$ ancilla qubits are measured in
the computational basis. In the absence of decoherence ($t=0$) the
state after decoding is $(\alpha |0\rangle +\beta |1\rangle) \,
|0^{N-1}\rangle$, so that the initial state of the main qubit is
restored and the measurement results for all ancillas are 0. The
decoherence disturbs the final state, which probabilistically
changes the measurement results and the corresponding final states
of the main qubit.

\begin{figure}[ptb]
\centering
\includegraphics[width=3.0in]{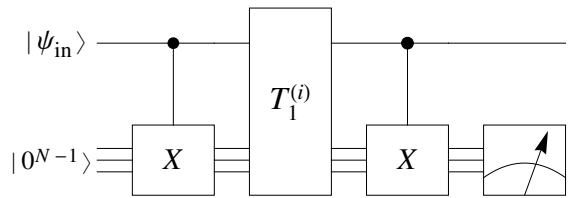}
\caption{$N$-qubit repetitive coding algorithm with one control
qubit initially containing the quantum information. The
controlled-$X$ block represents CNOT gates from the main qubit to
each ancilla qubit individually. $T_{1}^{(i)}$ represents energy
relaxation of the $i$th qubit ($i=1$ for the main qubit, $i\geq 2$
for ancilla qubits).
    } \label{T1 protocol}
\end{figure}

    Even when the measurement result is all $N-1$ zeros (for which
we will use the bold-font notation $\bf 0$), the state of the main
qubit is not exactly $|\psi_{\rm in}\rangle$; however, we will see
that it is close to  $|\psi_{\rm in}\rangle$. A measurement result
different from $\bf 0$ indicates an error. There are three ways to
handle this situation. First, the measurement result can be simply
ignored; in this case there is obviously no benefit from using the
encoding/decoding procedure. Second, we can reject such cases and
keep only realizations with the measurement result ${\bf 0}$; we
will refer to this selective procedure as quantum error detection.
Third, we can apply a quantum operation to the main qubit to make
its state closer to $|\psi_{\rm in}\rangle$. This operation will
depend on the measurement result, and the procedure is then quantum
error correction.

    For simplicity in this section we neglect decoherence
(and other imperfections) during encoding, decoding, and
measurement; it will be taken into account in the next section when
we will discuss realistic experiments with phase qubits.
    To characterize the efficiency of a procedure either the
quantum process tomography (QPT) fidelity $F_\chi$ or the average
state fidelity $F_{\rm av}$ can be used. The QPT fidelity is usually
defined as \cite{N-C,Nielsen-02} $F_\chi=\mbox{ Tr} (\chi_{\rm
desired}\chi )$, where $\chi$ is the process matrix and $\chi_{\rm
desired}$ in our case corresponds to the ideal quantum memory
operation, i.e.\ no evolution of the logic qubit. The average state
fidelity is \cite{N-C,Nielsen-02} $F_{\rm av} = \int \mbox{Tr}
(\rho_{\rm fin} U_0|\psi_{\rm in} \rangle \langle \psi_{\rm in} |
U_0^\dagger )\, d|\psi_{\rm in} \rangle $, where $U_0=\openone$ is
the desired unitary operator, $\rho_{\rm fin}(|\psi_{\rm
in}\rangle\langle \psi_{\rm in}|)$ is the actual mapping from the
initial state to the final density matrix $\rho_{\rm fin}$, and the
normalized integral is over all pure initial states $|\psi_{\rm
in}\rangle$ using the Haar measure. For a trace-preserving operation
$F_{\rm av}=(F_\chi d+1)/(d+1)$, where $d=2$ is the dimension of our
Hilbert space \cite{Nielsen-02}. This relation holds for QEC and/or
when the measurement result is ignored. However, for the QED
procedure there is a problem \cite{Keane} in defining the QPT
fidelity $F_\chi$ because the procedure is selective; then the
quantum operation for normalized states is not linear and the
corresponding (trace-preserving) matrix $\chi$ cannot be defined
rigorously. In this case we {\it define} $F_\chi$ via the average
state fidelity,
    \be
F_{\chi} = (3 F_{\rm av}-1)/2,
    \label{F-chi-av}\ee
as for a trace-preserving operation.

\subsection{Single-qubit relaxation}

    Before calculating the fidelity of the QEC and QED procedures, let
us calculate the quantum memory fidelity of a single qubit, without
any encoding. We also consider first this simple case to demonstrate
a technique of unraveling the evolution due to energy relaxation,
which is later used for $N$-qubit encoding.

    After time $t$ an initial state
$|\psi_{\rm in}\rangle= \alpha |0\rangle +\beta |1\rangle$ becomes a
density matrix (here the upper row and left column correspond to the
excited state $|1\rangle$)
    \be
\rho_{\rm fin} = \left(\begin{array}{cc} |\beta |^2 e^{-t/T_1}
& \alpha^* \beta e^{-t/2T_1} \\
\alpha \beta^* e^{-t/2T_1}& |\alpha |^2 +|\beta |^2 (1-e^{t/T_1})
    \end{array}\right) ,
    \ee
which can be represented using the Kraus operators:
    \begin{eqnarray}
&& \rho_{\rm fin} = A_{\rm r} \rho_{\rm in} A_{\rm r}^\dagger +
A_{\rm n} \rho_{\rm in} A_{\rm n}^\dagger ,
   \label{unravel-1} \\
&&  A_{\rm r} =  \left(\begin{array}{cc} 0
& 0 \\
\sqrt{p}&  0 \end{array}\right), \,\,\,
 A_{\rm n} = \left(\begin{array}{cc} \sqrt{1-p}
& 0 \\
0&  1 \end{array}\right),
    \label{Kraus}\end{eqnarray}
where $p=1-e^{-t/T_1}$, $\rho_{\rm in}=|\psi_{\rm in}\rangle \langle
\psi_{\rm in} |$, and the Kraus operators satisfy the completeness
relation $ A_{\rm r}^\dagger A_{\rm r}+ A_{\rm n}^\dagger A_{\rm
n}=\openone$. This representation has an obvious interpretation as
two scenarios of the evolution. The first term in Eq.\
(\ref{unravel-1}) corresponds to qubit relaxation into the ground
state $|0\rangle$ with probability $P_{\rm r}=|\beta|^2 p$. The
second term is the no-relaxation scenario, which occurs with the
remaining probability $P_{\rm n}=|\alpha|^2+|\beta |^2
(1-p)=1-P_{\rm r}$ and transforms the qubit into the state
    \be
    |\psi_{\rm n}\rangle =\frac{A_{\rm n} |\psi_{\rm in}\rangle}
    {\sqrt{P_{\rm n}}} =
\frac{\alpha |0\rangle +\beta \sqrt{1-p}
    \,
     |1\rangle}{\sqrt{P_{\rm n}}}.
    \ee
The non-unitary evolution $|\psi_{\rm in}\rangle \rightarrow
|\psi_{\rm n}\rangle$  is essentially the same as for a partial
collapse due to a null-result measurement in the experiment of Ref.\
\cite{Katz-Science}.

    Now let us find the averaged state fidelity $F_{\rm av}=\overline{
\mbox{Tr}(\rho_{\rm f} |\psi_{\rm in}\rangle \langle \psi_{\rm in}
|) }$ using unraveling into the relaxation and no-relaxation
scenarios. With probability $P_{\rm r}$ the state fidelity is
$F_{\rm st,r}=|\langle 0|\psi_{\rm in}\rangle |^2=|\alpha|^2$, and
with probability $P_{\rm n}$ the state fidelity is $F_{\rm
st,n}=|\langle \psi_{\rm n} |\psi_{\rm in}\rangle
|^2=(|\alpha|^2+\sqrt{1-p}|\beta|^2)^2/P_{\rm n}$. Therefore for an
initial state $|\psi_{\rm in}\rangle$ the state fidelity is
    \begin{eqnarray}
&& \hspace{-0.3cm}    F_{\rm st} = F_{\rm st,r} P_{\rm r}+ F_{\rm
st,n} P_{\rm n}
    \label{F-st-1}\\
&&=     |\alpha|^2 |\beta |^2 p + |\alpha|^4 +(1-p)
    |\beta|^4 +2|\alpha|^2 |\beta|^2\sqrt{1-p}, \qquad
    \label{F-st}
    \end{eqnarray}
and the average fidelity $F_{\rm av}=\overline {F_{\rm st}}$ can be
calculated by averaging $|\alpha|^4$, $|\beta|^4$, and
$|\alpha|^2|\beta|^2$ over the Bloch sphere. These averages
(including some others for completeness and later use) are
    \begin{eqnarray}
&&    \overline{|\alpha|^4}=\overline{|\beta|^4} =
 \int_0^\pi \frac{(1+\cos\theta )^2 }{4} \, \frac{\sin \theta}{2}\,
       d\theta =\frac{1}{3} , \qquad
        \label{Bloch-av-1}   \\
&&     \overline{|\alpha|^2|\beta|^2}=
 \int_0^\pi \frac{(1+\cos\theta )(1-\cos\theta) }{4} \, \frac{\sin \theta}{2}\,
       d\theta =\frac{1}{6} , \qquad
   \label{Bloch-av-2} \\
&&    \overline{|\alpha|^2}=\overline{|\beta|^2} =
       \frac{1}{2} , \,\,\,
       \overline{|\alpha|^6}=\overline{|\beta|^6} = \frac{1}{4} ,
        \label{Bloch-av-3} \\
&&  \overline{|\alpha|^2|\beta|^4}=
\overline{|\alpha|^4|\beta|^2}=\frac{1}{12} , \,\,\,
   \overline{|\alpha|^4|\beta|^4}= \frac{1}{30} , \qquad
    \label{Bloch-av-4} \\
&& \overline{1/(A+B|\beta|^2)}= (1/B) \ln (1+B/A), \qquad
        \label{Bloch-av-5}   \\
&& \overline{|\beta|^2/(A+B|\beta|^2)}= (1/B) -(A/B^2) \ln (1+B/A),
\qquad
        \label{Bloch-av-6}   \\
&& \overline{|\beta|^4/(A+B|\beta|^2)}=
\frac{1}{2B}-\frac{A}{B^2}+\frac{A^2}{B^3} \ln (1+B/A),
 \qquad
   \label{Bloch-av-7} \\
&& \overline{\frac{|\alpha|^4}{A+B|\beta|^2}}=
\frac{-3}{2B}-\frac{A}{B^2}+\frac{(A+B)^2}{B^3} \ln (1+B/A),
 \qquad
   \label{Bloch-av-7a} \\
   && \overline{\frac{|\alpha|^2|\beta|^2}{A+B|\beta|^2}}=
   \frac{1}{2B}+\frac{A}{B^2}-\frac{A(A+B)}{B^3} \ln (1+B/A),
 \qquad
   \label{Bloch-av-8}
    \end{eqnarray}
where $A$ and $B$ are constants, and we used integration over the
Bloch-sphere polar angle $\theta$, so that $|\alpha|^2=(1+\cos
\theta )/2$ and $|\beta|^2=(1-\cos \theta )/2$.

Applying the averages (\ref{Bloch-av-1}) and (\ref{Bloch-av-2}) to
Eq.\ (\ref{F-st}), we obtain the average state fidelity
    \be
    F_{\rm av} = \frac{2}{3} +\frac{\sqrt{1-p}}{3} -\frac{p}{6}.
    \label{1q-av}\ee Actually, there is an easier way to obtain
this result. Instead of averaging $F_{\rm st}$ over the Bloch
sphere, it is sufficient \cite{Nielsen-02,Bowdrey-02} (see also
Appendix) to calculate the average only over 6 initial states:
$|0\rangle$, $|1\rangle$, $(|0\rangle\pm |1\rangle)/\sqrt{2}$, and
$(|0\rangle\pm i |1\rangle)/\sqrt{2}$. However, in our further
analysis this trick does not always help, so we prefer the full
integration over the Bloch sphere. Using Eq.\ (\ref{F-chi-av}) it is
easy to convert Eq.\ (\ref{1q-av}) into the QPT fidelity: $F_{\chi}=
(1+\sqrt{1-p}-p/2)/2$. Note that for small $p$
    \be
      F_{\rm av} \approx 1 - \frac{p}{3}, \,\,\,
    F_{\chi} \approx 1 - \frac{p}{2}, \,\,\, p\approx \frac{t}{T_1} \ll 1.
    \label{1q-av-2}\ee

   The average state fidelity (\ref{1q-av}) is averaged over the two
scenarios. Let us now discuss the average state fidelity in each
scenario separately, having in mind a gedanken experiment in which
an emitted photon or phonon is always captured and recorded, thus
allowing us to distinguish the two scenarios. If the relaxation has
happened, then $F_{\rm st,r}=|\alpha|^2$ and averaging this over the
Bloch sphere we obtain
    \be
F_{\rm av, r}=\overline{|\alpha|^2}=1/2.
    \ee
 Similarly, for the
no-relaxation scenario $F_{\rm av,n}=
\overline{(|\alpha|^2+\sqrt{1-p}|\beta|^2)^2/
[|\alpha|^2+(1-p)|\beta|^2]}$, which can be calculated using Eqs.\
(\ref{Bloch-av-7})--(\ref{Bloch-av-8}):
 \begin{eqnarray}
&& F_{\rm av, n}= \frac{1}{2}+ \frac{\sqrt{1-p}(2-p)-2 (1-p) }{p^2}
    \nonumber\\
&&\hspace{0.9cm} +\frac{(1-p) (2 \sqrt{1-p}-2+p)}{p^3} \,  \ln(1-p)
\qquad
    \end{eqnarray}
For $p\ll 1$ this gives $F_{\rm av, n}\approx 1- p^2/24$, showing a
slow, quadratic in time decrease of fidelity in the no-relaxation
scenario in contrast to the linear decrease (\ref{1q-av-2}) of the
fidelity averaged over both scenarios. Therefore our gedanken
experiment could be used for quantum error detection: if no
relaxation is recorded, we know that the initial state is
well-preserved at short times.

    Note that we have averaged the state fidelities $F_{\rm st,r}$
and $F_{\rm st,n}$ over the Bloch sphere with uniform weight, as in
the standard definition \cite{N-C,Nielsen-02} of the averaged state
fidelity. Another meaningful averaging is using weights proportional
to the probabilities of the corresponding scenarios. [This would
correspond to an equal number of experimental runs for each point of
a uniform mesh on the Bloch sphere, as opposed to an equal number of
``successful'' (i.e.\ selected) runs for the previous definition.]
Thus defined average fidelities are
    \begin{eqnarray}
\widetilde F_{\rm av, r}=\overline{|\alpha|^2P_{\rm
r}}/\overline{P_{\rm r}}=1/3, \qquad\qquad\qquad
    \\
\widetilde F_{\rm av,n}=
\overline{(|\alpha|^2+\sqrt{1-p}|\beta|^2)^2}/\overline{P_{\rm
n}}=\frac{2-p+\sqrt{1-p}}{3-3p/2}, \quad
    \end{eqnarray}
 where $\overline{P_{\rm r}}=p/2$ and
$\overline{P_{\rm n}}=1-p/2$ are the averaged probabilities of the
two scenarios. The advantage of this definition is a natural formula
for the non-selected average fidelity:
    \be
    F_{\rm av}=  \widetilde F_{\rm av,r} \overline{P_{\rm r}} +
     \widetilde F_{\rm av,n} \overline{P_{\rm n}}
        \label{F-av-tilde-F}\ee
[see Eq.\ (\ref{F-st-1})]. In this paper when discussing selected
scenarios (as for QED) we will use both ways to average over the
Bloch sphere. Note that $\widetilde F_{\rm av,n} \approx 1- p^2/24$
for $p\ll 1$, which is the same as for $F_{\rm av,n}$ ($\widetilde
F_{\rm av,n}$ and $F_{\rm av,n}$ are practically indistinguishable
at $p\alt 1/2$), indicating that the difference between the two
definitions is not very significant in the cases that are of most
interest for this paper.

\subsection{Two-qubit encoding}

    Let us use the procedure of Fig.\ 1 with only one ancilla qubit.
The encoded state is then $\alpha |00\rangle +\beta |11\rangle$. The
state evolution due to energy relaxation can be unraveled into four
scenarios: no relaxation, relaxation in either the first (main) or
second (ancilla) qubit, and relaxation in both qubits. The
corresponding wavefunctions and probabilities after time $t$ of
energy relaxation are
   \begin{equation}
    \left\{ \begin{array}{llll} \displaystyle
    \frac{\alpha|00\rangle+\beta\sqrt{1-p_1}\sqrt{1-p_2}|11\rangle}
    {\sqrt{P_{\rm nn}}},
     \\
    \hspace{0.9cm} \mbox{prob.} \,\,\, P_{\rm nn}=  \abs{\alpha}^2+
    \abs{\beta}^2     (1-p_1)(1-p_2),
        \\
    |01\rangle, \,\,\, \mbox{prob.} \,\,\,
    P_{\rm rn}=\abs{\beta}^2 p_1(1- p_2),\\
    |10\rangle, \,\,\, \mbox{prob.} \,\,\,
    P_{\rm nr}=\abs{\beta}^2 (1- p_1)p_2,\\
    |00\rangle, \,\,\, \mbox{prob.} \,\,\,
    P_{\rm rr}=\abs{\beta}^2  p_1 p_2,
    \end{array} \right.
    \label{2q-unrav}\end{equation}
where
    \be
p_1=1-e^{-t/T_{1}^{(1)}}, \,\,\, p_2=1-e^{-t/T_{1}^{(2)}}
    \ee
 are the
single-qubit probabilities of relaxation from the excited state
$|1\rangle$. This simple unraveling is possible because the energy
relaxation occurs only in component $|11\rangle$ of the
superposition, and in this component the qubits are unentangled.
This is why the probabilities of the scenarios are the simple
products of individual probabilities. The validity of Eq.\
(\ref{2q-unrav}) can also be checked by considering particular time
moments at which the relaxation events happen and integrating over
these moments; this is a more direct but more cumbersome way.

    After the decoding procedure consisting of one CNOT operation,
the two-qubit state is a product-state in all four scenarios:
  \begin{equation}
    \left\{ \begin{array}{llll} \displaystyle
    \frac{\alpha|0\rangle+\beta\sqrt{1-p_1}\sqrt{1-p_2}|1\rangle}
    {\sqrt{P_{\rm nn}}} \otimes |0\rangle , \, \mbox{prob.} \,\,\, P_{\rm nn},
        \\
    |01\rangle, \,\,\, \mbox{prob.} \,\,\,
    P_{\rm rn},\\
    |11\rangle, \,\,\, \mbox{prob.} \,\,\,
    P_{\rm nr},\\
    |00\rangle, \,\,\, \mbox{prob.} \,\,\,
    P_{\rm rr},
    \end{array} \right.
    \label{2q-unrav-2}\end{equation}
with a definite result of the ancilla qubit measurement in each
scenario. The state of the main qubit is different from the initial
state $|\psi_{\rm in}\rangle$ in all four scenarios, and the
corresponding state fidelities are $[|\alpha|^2+|\beta|^2
\sqrt{1-p_1}\sqrt{1-p_2}]^2/P_{\rm nn}$,  $|\alpha|^2$, $|\beta|^2$,
and $|\alpha|^2$.

    As discussed above, we consider three ways to proceed: ignore
the measurement result, select only result 0, or try to correct the
main qubit state. If the measurement result is ignored, then all
four scenarios are added up and the average fidelity is
    \begin{eqnarray}
&&    F_{\rm av}^{\rm ign} = \overline{[|\alpha|^2+|\beta|^2
\sqrt{1-p_1}\sqrt{1-p_2}]^2}
    \nonumber \\
&&\hspace{1cm} + \overline{|\alpha|^2 P_{\rm rn} +|\beta|^2
P_{\rm nr} +|\alpha|^2 P_{\rm rr}} ,
    \end{eqnarray}
where the averaging is over the Bloch sphere.
 Using the formulas for the probabilities from Eq.\ (\ref{2q-unrav})
and the averages $\overline{|\alpha|^2}$, $\overline{|\beta|^2}$,
$\overline{|\alpha|^4}$,  $\overline{|\beta|^4}$,   and
$\overline{|\alpha|^2 |\beta|^2}$ from Eqs.\
(\ref{Bloch-av-1})--(\ref{Bloch-av-3}), we obtain
   \be
   F_{\rm av}^{\rm ign} =
   \frac{2}{3}+\frac{\sqrt{(1-p_1)(1-p_2)}}{3}-\frac{p_1}{6}  \,\,
    \label{2q-av-ign}\ee
For small $p_{1,2}$ (at short time $t$) it is  $F_{\rm av}^{\rm ign}
\approx 1- p_1/3 -p_2/6$, and it is obviously worse than the case
without encoding/decoding of the main qubit -- see Eqs.\
(\ref{1q-av}) and (\ref{1q-av-2}). Note that Eq.\ (\ref{2q-av-ign})
can also be obtained by averaging the state fidelity only over the 6
initial states (see Appendix).

    In quantum error detection we consider ancilla measurement result
1 as an error and select only the cases when the measurement gives
0. This selects scenarios with either no relaxation or two
relaxation events [see the first and last lines of Eq.\
(\ref{2q-unrav-2})]. The averaged (with uniform weight) state
fidelity in this case is
    \be
  F_{\rm av}^{\rm qed} =  \overline{\frac{[|\alpha|^2+|\beta|^2
\sqrt{1-p_1}\sqrt{1-p_2}]^2 +|\alpha|^2 P_{\rm rr}}{P_{\rm
nn}+P_{\rm rr}} }
    \ee
(the fraction is averaged over the Bloch sphere), which can be
calculated using Eqs.\ (\ref{Bloch-av-7})--(\ref{Bloch-av-8}):
    \begin{eqnarray}
&& F_{\rm av}^{\rm qed}= \frac{1}{2}+\frac{s-1}{B}+
\frac{p_1+p_2-2+2s}{B^2}
    \nonumber \\
&& \hspace{0.5cm} +\frac{(1+B)^2+s^2-(1+B)(2s+p_1p_2)}{B^3}
\ln(1+B), \qquad
   \label{2q-qed}
    \end{eqnarray}
where $B=2p_1p_2-p_1-p_2$ and $s=\sqrt{1-p_{1}}\sqrt{1-p_{2}}$. For
the small-error case (at short time $t$) this gives
    \be
F_{\rm av}^{\rm qed}\approx 1-(
p_{1}^{2}+p_{2}^{2})/24-5p_{1}p_{2}/12, \,\,\, p_{1,2}\ll 1.
    \label{2q-av-qed-approx}\ee

    If we use the averaging over the Bloch sphere with weight proportional
to the probability $P_{\rm nn}+P_{\rm rr}$ of the measurement result
0, then
   \be
  \widetilde  F_{\rm av}^{\rm qed} =  \frac{\overline{[|\alpha|^2+|\beta|^2
\sqrt{1-p_1}\sqrt{1-p_2}]^2 +|\alpha|^2 P_{\rm
rr}}}{\overline{P_{\rm nn}+P_{\rm rr}} } ,
    \label{2q-av-tilde}\ee
(the numerator and denominator are averaged separately), which gives
  \be
  \widetilde  F_{\rm av}^{\rm qed} =
    \frac{2-p_1-p_2 + \frac{3}{2}p_1p_2
    +\sqrt{1-p_1}\sqrt{1-p_2}}{3[1+p_1p_2-(p_1+p_2)/2]} .
  \label{2q-av-t2}\ee
  [Note that instead of using Eqs.\ (\ref{Bloch-av-1})--(\ref{Bloch-av-3}),
the 6-point averaging trick can be used separately for the numerator
and denominator of Eq.\ (\ref{2q-av-tilde}) -- see Appendix.] At
short times this gives $\widetilde F_{\rm av}^{\rm qed} \approx 1-
(p_{1}^{2}+p_{2}^{2})/24 - 5p_{1}p_{2}/12$, same as for $F_{\rm
av}^{\rm qed}$ [see Eq.\ (\ref{2q-av-qed-approx})].

    Figure 2 shows the QED fidelity defined in both ways,
$ F_{\rm av}^{\rm qed}$ and $ \widetilde F_{\rm av}^{\rm qed}$, as
functions of the one-qubit relaxation probability, assuming similar
qubits, $p_1=p_2=p$. For $p\alt 0.3$ both fidelities are
significantly higher than the fidelity $F_{\rm av}^{\rm 1q}$ for an
unencoded single qubit [given by Eq.\ (\ref{1q-av})], which itself
is higher than the fidelity $F_{\rm av}^{\rm ign}$ when the ancilla
measurement result is ignored [Eq.\ (\ref{2q-av-ign})].

\begin{figure}[ptb]
\centering
\includegraphics[width=3.2in]{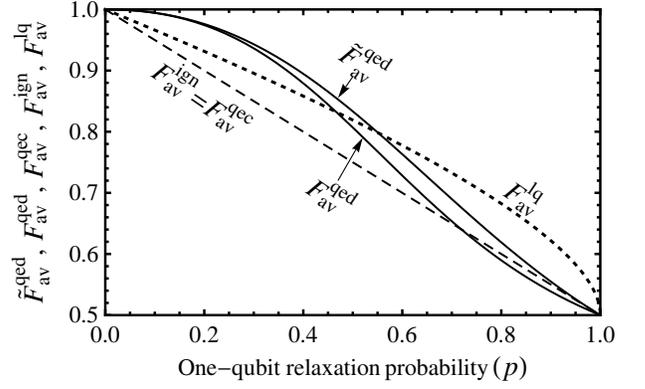} \caption{Average state
preservation fidelities for the two-qubit encoding (compared with no
encoding), as functions of the one-qubit energy relaxation
probability $p=1-e^{-t/T_1}$ (same for both qubits, $p_1=p_2=p$).
The solid lines show the QED fidelities $F_{\rm av}^{\rm qed}$ and
$\widetilde F_{\rm av}^{\rm qed}$ given by Eqs.\ (\ref{2q-qed}) and
(\ref{2q-av-t2}). ($F_{\rm av}^{\rm qed}$ assumes averaging over the
Bloch sphere with uniform weight, while for $\widetilde F_{\rm
av}^{\rm qed}$ the weight is proportional to the probability of the
``no error'' measurement result 0.) The dashed line shows the QEC
fidelity $F_{\rm av}^{\rm qec}$, which coincides with $F_{\rm
av}^{\rm ign}$, for which the measurement result is ignored, Eqs.\
(\ref{2q-av-ign}) and (\ref{2q-qec-2}). The dotted line shows the
one-qubit fidelity $F_{\rm av}^{\rm 1q}$ without encoding, Eq.\
(\ref{1q-av}).
  QEC performs worse than no encoding, while QED provides a
significant improvement for $p\alt 0.3$.
   } \label{two qubit T1 graph}
\end{figure}

   Now let us discuss whether or not the state of the main qubit can
be made closer to $|\psi_{\rm in}\rangle$ using the measurement
result information, as in quantum error correction. If the
measurement result is 0, then the qubit state is described by the
first and last lines of Eq.\ (\ref{2q-unrav-2}). It is rather
obvious that in this case no unitary operation can improve further
the average fidelity [for QEC we are interested in averaging with
the weight proportional to probability -- see Eq.\
(\ref{F-av-tilde-F})]. This statement is rigorously proven in the
Appendix. So, no correction should be applied for measurement result
0. (Actually, a non-unitary operation due to partial collapse can
increase the fidelity in this case
\cite{Keane,Katz-Science,Zubairy,Kim}, but we consider only unitary
operations, as it should be in the usual QEC.)
 When the measurement result is 1, the main qubit is in the state
$|0\rangle$ with probability $P_{\rm rn}/(P_{\rm rn}+P_{\rm nr})$ or
in the state $|1\rangle$ with remaining probability $P_{\rm
nr}/(P_{\rm rn}+P_{\rm nr})$ -- see Eq.\ (\ref{2q-unrav-2}). In the
case $p_1=p_2$ this is the fully mixed state, and any unitary
operation does not change it. Thus a meaningful error correction is
impossible, and therefore
    $F_{\rm av}^{\rm qec}=F_{\rm av}^{\rm ign}$ (see Fig.\ 2).

Actually, if $p_2> p_1$, then a slight improvement of fidelity is
possible by applying the $\pi$-pulse (exchanging states $|0\rangle$
and $|1\rangle$) when the measurement result is 1. This makes the
resulting state closer to $|1\rangle$ than to $|0\rangle$, and
correspondingly on average closer to $|\psi_{\rm in}\rangle$,
because the probability of measuring 1 increases with $|\psi_{\rm
in}\rangle$ being closer to $|1\rangle$. The optimality of this
procedure for measurement result 1 is proven in the Appendix. It is
easy to calculate the fidelity change due to the $\pi$-pulse (the
easiest way is to average over the 6 initial states and to work with
unnormalized states -- see Appendix). The resulting optimal QEC
fidelity for the 2-qubit encoding of Fig.\ 1 is
   \be
   F_{\rm av}^{\rm qec} =
   \frac{2}{3}+\frac{\sqrt{(1-p_1)(1-p_2)}}{3}-\frac{\min
   (p_1,p_2)}{6}.
    \label{2q-qec-2}\ee

\subsection{$N$-qubit encoding}

We now extend our discussion of the protocol of Fig.\ \ref{T1
protocol}, including $N-1$ ancilla qubits. The encoded state is then
$\alpha|0^{N}\rangle +\beta|1^{N}\rangle$. The state evolution can
be unraveled into $2^{N}$ scenarios depending on which qubits relax.
However, there are $2^{N-1}$ measurement results, and each of them
corresponds to two scenarios. If the main qubit does not relax, then
the measurement result directly shows which ancilla qubits relax
(i.e.\ result 1 indicates the relaxation event), while if the main
qubit relaxes, then the relaxation scenario is shown by the
complement of the measurement result (i.e.\ result 0 indicates
relaxation).

    The measurement result $\bf 0$ (all zeros) indicates that the main
qubit is either in the state
\begin{equation}
|\psi_{\rm none}\rangle= \frac{1}{\sqrt{P_{\rm none}}}
\left(\alpha|0\rangle
+\beta|1\rangle\prod\nolimits_{j=1}^{N}\sqrt{1-p_{j}}\right)  ,
    \label{Nq-none}
\end{equation}
where $P_{\rm none}=|\alpha|^{2}+|\beta|^{2}
{\textstyle\prod_{j=1}^{N}}(1-p_{j})$ is the probability that no
qubits relax, or in the state $|0\rangle$ if all qubits relax, with
the corresponding probability $P_{\rm all}=|\beta|^{2}
{\textstyle\prod_{j=1}^{N}} p_{j}$. For any other measurement result
the main qubit is either in state $|0\rangle$ or $|1\rangle$, with
easily calculable probabilities of the scenarios. For simplicity we
assume $p_j=p$ below.

    As in the previous subsection, we consider three possible ways
to proceed: ignore the measurement result, select only cases with
measurement result $\bf 0$ (QED), or try to improve the fidelity
when an error is detected (QEC). If the measurement result is
ignored, the average fidelity (calculated in a similar way as above)
is
\begin{equation}
F_{\rm av}^{\rm ign}=\frac{2}{3}+\frac{(1-p)^{N/2}}{3}-\frac{p}{6};
\end{equation}
it obviously decreases with increasing number of ancilla qubits.

In quantum error detection we select only cases with measurement
result $\bf 0$. Then the state fidelity is
    \be
F_{\rm st}^{\rm qed}=
\frac{(|\alpha|^{2}+|\beta|^{2}(1-p)^{N/2})^{2}+ |\alpha|^{2}P_{\rm
all}}{P_{\rm none}+P_{\rm all}},
    \label{Nq-qed-state}\ee
     and averaging
it over the Bloch sphere with uniform weight we obtain
\begin{eqnarray}
&& \hspace{-0.5cm} F_{\rm av}^{\rm qed}=\frac{-3+S+(1-p)^{N}}{2B}
+\frac{-1+S-(1-p)^{N}}{B^2}
    \nonumber \\
&&+\frac{(1+B)^{2}+(1-p)^{N}-S(1+B)}{ B^3}\, \ln(1+B), \qquad
   \end{eqnarray}
where $B=-1+(1-p)^{N}+p^{N}$ and $S=2(1-p)^{N/2}+p^N$. For $N=2$
this equation corresponds to Eq.\ (\ref{2q-qed}). The small-error
approximation for $N\geq 3$ is
    \be
F_{\rm av}^{\rm qed}\approx1-N^{2}p^{2}/24, \,\,\, p\ll 1.
    \label{Nq-qed-approx}\ee
It is interesting to note that this approximation does not work for
$N=2$, for which $F_{\rm av}^{\rm qed}\approx 1- p^{2}/2$, as
follows from Eq.\ (\ref{2q-av-qed-approx}). The reason is that
$P_{\rm all}$ scales as $p^N$, and therefore for $N\geq 3$ Eq.\
(\ref{Nq-qed-approx}) does not have a quadratic contribution from
the scenario when all qubits relax; the infidelity comes only from
the difference between $|\psi_{\rm none}\rangle$ and $|\psi_{\rm
in}\rangle$. In contrast, for $N=2$ the fidelity $F_{\rm av}^{\rm
qed}$ is further decreased by $p^2/3$ due to relaxation of both
qubits.

    From Eqs.\ (\ref{Nq-qed-approx}) and (\ref{2q-av-qed-approx}) we
see that the best QED fidelity in the small-error case ($p\ll 1$) is
achieved by the 3-qubit encoding, $N=3$; then $1-F_{\rm av}^{\rm
qed}\approx (3/8)\, p^2$. However, this is only the factor $4/3$
better (smaller) than for $N=2$. Therefore, from the experimental
point of view the 2-qubit encoding (which is easier to realize than
the 3-qubit encoding) seems to be most natural.

If we use the averaging of the QED state fidelity
(\ref{Nq-qed-state}) with weights proportional to the probability
$P_{\rm none}+P_{\rm all}$ of the measurement result $\bf 0$, then
we essentially  need to average the numerator and denominator of
Eq.\ (\ref{Nq-qed-state}) separately, thus obtaining
\begin{equation}
\widetilde{F}_{\rm av}^{\rm qed}=\frac{2}{3}\,
\frac{1+(1-p)^{N}+(1-p)^{N/2}+ \frac{1}{2}p^{N}}{1+(1-p)^{N}+p^{N}}.
\end{equation}
In the small-error case ($p\ll 1$) for $N\geq 3$ this gives
 $\widetilde{F}_{N}^{\rm qed} \approx1-N^{2}p^{2}/24,$ same
as Eq.\ (\ref{Nq-qed-approx}) for $F_{N}^{\rm qed}$.

Figure 3(a) shows the QED fidelities $F_{\rm av}^{\rm qed}$ and
$\widetilde{F}_{\rm av}^{\rm qed}$ for $N=2$, 3, and 4. We see that
the difference between $F_{\rm av}^{\rm qed}$ and
$\widetilde{F}_{\rm av}^{\rm qed}$ becomes larger with increasing
$N$, but the difference is small at small $p$. Note that the QED
fidelity for the 2-qubit encoding becomes better than for the
3-qubit encoding for $p\agt 0.3$.

\begin{figure}[ptb]
\centering
\includegraphics[width=3.1in]{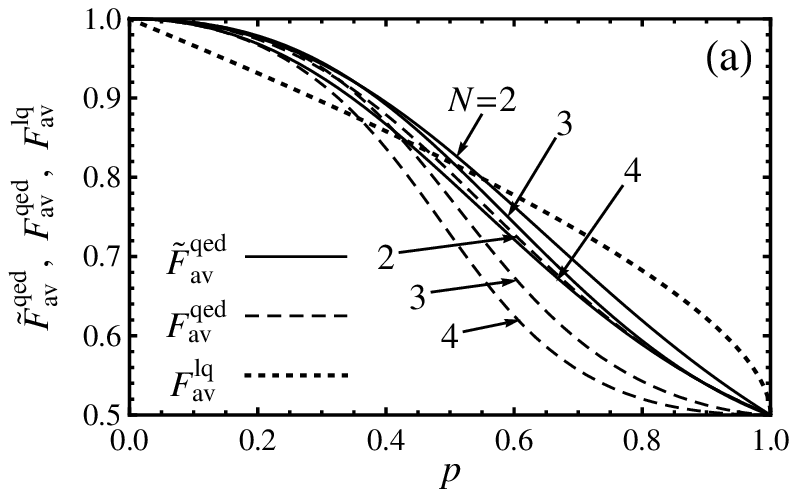}
\includegraphics[width=3.1in]{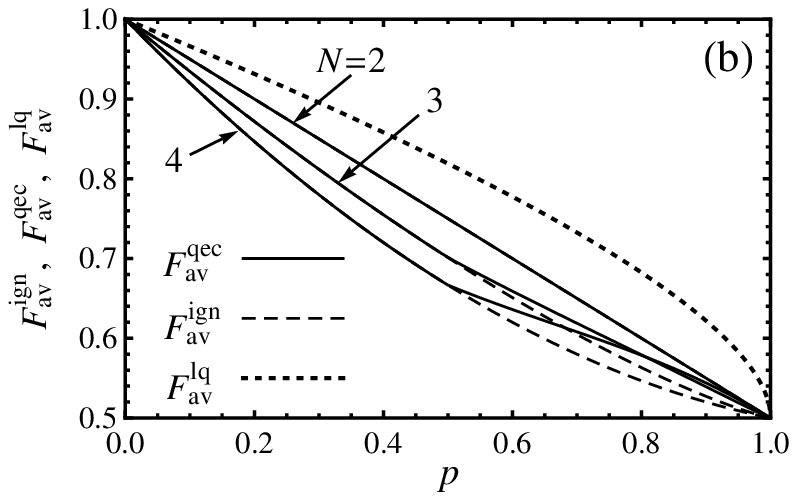}
    \caption{(a) The QED fidelities $\widetilde{F}_{\rm av}^{\rm qed}$
(solid lines) and  $F_{\rm av}^{\rm qed}$ (dashed lines) for the
encoding using $N=2$, 3, and 4 physical qubits, as functions of the
single-qubit energy relaxation probability $p$. The dotted line
shows the fidelity $F_{\rm av}^{\rm 1q}$ for an unencoded qubit.
    (b) The optimal QEC fidelity $F_{\rm av}^{\rm qec}$ (solid lines) and the
    fidelity $F_{\rm av}^{\rm ign}$ when the measurement result is
    ignored (dashed lines) for $N=2$, 3, and 4.
    }
\label{N qubit T1 graph}
\end{figure}

Now let us discuss the possibility of QEC protocols, which use
unitary correcting operations depending on the measurement result.
If the result  is $\bf 0$, then the unnormalized density matrix of
the main qubit is
  $P_{\rm
none}|\psi_{\rm none}\rangle\langle\psi_{\rm none}|+ P_{\rm
all}|0\rangle\langle0|$. As proven in the Appendix, no unitary
operation can increase the fidelity in this case (in contrast to
non-unitary partial-collapse operations
\cite{Keane,Katz-Science,Zubairy,Kim}). For all other measurement
results, the main qubit is in the incoherent mixture of the states
$|0\rangle$ and $|1\rangle$; the unnormalized density matrix is $
P_{0,{\rm m}}|0\rangle\langle 0|+ P_{1,{\rm m}}|1\rangle\langle 1|$,
where the corresponding probabilities are $P_{1,{\rm m}}= |\beta|^2
(1-p)\prod_{i=2,N}f(m_i)$ and $P_{0,{\rm m}}= |\beta|^2
p\prod_{i=2,N}f(1-m_i)$, where $f(1)=p$, $f(0)=1-p$, and $m_i$ is
the measurement result for the $i$th ancilla qubit. As shown in the
Appendix, the maximum fidelity is then achieved by applying the
$\pi$-pulse (exchanging $|0\rangle$ and $|1\rangle$) if $P_{1,{\rm
m}}<P_{0,{\rm m}}$ and doing nothing if $P_{1,{\rm m}}\geq P_{0,{\rm
m}}$. Calculating the corresponding qubit state fidelity (compared
with the initial state), summing over the $2^{N-1}$ measurement
results, and averaging over the Bloch sphere, we obtain the QEC
fidelity
\begin{eqnarray}
&& F_{\rm av}^{\rm qec}= \frac{1}{2} + \frac{1}{3} (1-p)^{N/2}+
\frac{1}{6}(1-p)^{N}
    \nonumber\\
&& \hspace{0.5cm} +\frac{1}{6} \max [p-p^{N},\, (1-p)-(1-p)^{N} ] .
\quad
\end{eqnarray}

The QEC fidelity as well as the fidelity $F_{\rm av}^{\rm ign}$ for
ignoring the measurement result are shown in Fig.\ 3(b) for $N=2$,
3, and 4. The curves for $F_{\rm av}^{\rm qec}$ and $F_{\rm av}^{\rm
ign}$ coincide at $p\leq 1/2$, because in this case the optimal
correction is no correction. We see that for any $N$ the QEC
fidelity is smaller than the no-encoding fidelity $F_{\rm av}^{\rm
1q}$, so the error correction by a repetitive code does not protect
against energy relaxation.

\subsection{Discussion}

    Our results show that the repetitive codes do not work for QEC
protection against energy relaxation. This is because energy
relaxation is very different from a bit flip (or phase flip or
bit-phase flip), for which repetitive codes work well.
  In the language of quantum stabilizer codes \cite{N-C}
the event of energy relaxation corresponds to the ``sum'' of two
errors: bit flip ($X$-operation) and bit-phase flip ($Y$-operation)
-- see the Kraus operator $A_{\rm r}$ in Eq.\ (\ref{Kraus}).
    So, a stabilizer code should be able to protect against both of
these errors to protect against energy relaxation events. [Actually,
a weaker error due to the ``no relaxation'' Kraus operator $A_{\rm
r}$ in Eq.\ (\ref{Kraus}) also requires protection against phase
flip ($Z$-operation) errors.]
    For example, the standard
5-qubit and 7-qubit QEC codes \cite{QEC-theory,N-C,Laflamme-96}
protect against all 3 types of errors ($X,Y,Z$), and therefore can
protect against energy relaxation.

    Using the approach of the stabilizer codes and the quantum
Hamming bound \cite{N-C}, let us calculate the minimum number of
qubits $N$ to protect against $X$ and $Y$ errors.
    The Hilbert space of the dimension $2^N$ can be divided into
$2^{N-1}$ orthogonal two-dimensional subspaces (``copies'' of the
qubit space); these subspaces should be able to distinguish the
cases with various errors and no error. Since the number of possible
errors is $2N$, we have an inequality $2^{N-1}\geq 1+2N$. From this
inequality we find $N\geq 5$ (the same minimum as for all 3 types of
quantum errors).
    Notice, however, that an approximate QEC for energy
relaxation is possible for $N=4$ \cite {Leung-4qubit} (see also
\cite{Shor-ampl-damping}). This code breaks the above limitation
because the relaxation event is treated as one error, not as a
``sum'' of $X$ and $Y$ (the drawback though is a slightly
probabilistic operation).
    In any case, the QEC codes protecting against energy
relaxation are much more complicated than the repetitive codes.

    Even though the repetitive codes are not good for QEC protection
against energy relaxation, we have shown that they can be well used
for QED. Moreover, only 2-qubit encoding is sufficient for that. An
interesting question is whether or not it is beneficial to do many
cycles of QED, correspondingly decreasing the time of each cycle and
therefore the error probability $p$ in each cycle (such division
into shorter cycles is beneficial for QEC \cite{QEC-theory,N-C}).
The simple answer is that such division into shorter cycles does not
help much for the protocol of Fig.\ 1. The reason is that even when
no relaxation events happen, the qubit state changes -- see Eq.\
(\ref{Nq-none}), because the absence of relaxation preferentially
indicates state $|0\rangle$ and plays the same role as the partial
collapse \cite{Katz-Science}. Rewriting Eq.\ (\ref{Nq-none}) in a
non-normalized way as $|\tilde \psi_{\rm none}\rangle=\alpha
|0\rangle + \beta |1\rangle \prod_{j=1}^N \exp (-t/2T_{1,j})$, we
see that division into several QED cycles does not change the final
wavefunction $|\tilde \psi_{\rm none}\rangle$ as long as the total
time $t$ is the same. Therefore, since for $N\geq 3$ this evolution
is the main reason for imperfect QED fidelity at $p\ll 1$ (see
discussion in the previous subsection), there is not much benefit of
using the QED cycles. Nevertheless, some improvement of the QED
fidelity will be due to a decrease of the probability $P_{\rm all}$
that all qubits relax. Since this probability scales as $p^N\approx
(t/T_1)^N$, the division into $M$ cycles is expected to decrease the
corresponding contribution to the procedure infidelity by the factor
$M^{N-1}$. This improvement will be most significant for $N=2$: it
will essentially change the approximation $F_{\rm av}^{\rm
qed}\approx 1- p^2/2$ given by Eq.\ (\ref{2q-av-qed-approx}) into
$F_{\rm av}^{\rm qed}\approx 1- p^2/6$ given by Eq.\
(\ref{Nq-qed-approx}) for $N=2$.

    A more important improvement of the QED fidelity can be achieved
if the no-relaxation evolution (\ref{Nq-none}) of $|\psi_{\rm
none}\rangle$ is compensated. One way is to apply a partial
measurement \cite{Keane,Katz-Science,Zubairy} to the main qubit
after the procedure, essentially eliminating the evolution
(\ref{Nq-none}) for the price of a further decrease of the selection
probability (probability of success). Another, easier way is to
apply $\pi$-pulses, exchanging states $|0\rangle$ and $|1\rangle$,
between (after) the QED cycles (these $\pi$-rotations can be around
any axis in the equatorial plane of the Bloch sphere). Then for an
even number of equal-duration QED cycles, the no-relaxation
evolution (\ref{Nq-none}) will be compensated exactly (as in the
uncollapsing procedure \cite{uncollapse,Keane,Zubairy,Kim}), and the
QED infidelity $1-F^{\rm qed}$ will be only due to the contribution
from $P_{\rm all}$. (The use of $\pi$-pulses resembles dynamical
decoupling of the ``bang-bang'' type \cite{bang-bang}; however, the
resemblance is accidental, since dynamical decoupling cannot protect
against the energy relaxation \cite{Pryadko}.)

For an estimate of the corresponding QED fidelity, let us consider
the procedure with total duration $t\alt T_1$, divided into $M$
cycles of duration $t/M$ each ($M$ is even). In each cycle $p\approx
t/M T_1 \ll 1$, and if we assume $N t/M T_1 \ll 1$, then $|\psi_{\rm
none}\rangle \approx |\psi_{\rm in}\rangle$ in Eq.\ (\ref{Nq-none}).
The probability that the $N$-qubit relaxation (which remains
undetected) happened in the first cycle is $P_{\rm all}=|\beta |^2
p^N$, and similar probability for the second cycle (after
$\pi$-pulse) is $|\alpha |^2 p^N$. Therefore, in a selected QED
realization (with all measurement results {\bf 0}) the probability
to have an undetected relaxation event is $(M/2)p^N$, independent of
the initial state (we assume this probability to be small, then we
can neglect the double-events). If such an undetected relaxation
event happens, then the average fidelity is $\widetilde F_{\rm
av}=\overline{|\alpha|^2|\beta|^2}/\overline{|\alpha|^2}=1/3$. The
QED fidelity then can be calculated as $1 -(1-1/3)\times (M/2)p^N$,
which gives
    \be
F_{\rm av}^{\rm qed}\approx \widetilde F_{\rm av}^{\rm qed}\approx
1-M(t/MT_1)^N /3.
    \label{Nq-cycles}\ee
(If the above assumption $N t/M T_1 \ll 1$ is violated, then the
factor $1/3$ changes, but the scaling remains the same.) We see that
for this procedure the division into a larger number of cycles $M$
is beneficial, as well as using more qubits ($N$) for the encoding.
    Note that our QED procedure does not prevent the relaxation
events from happening, so the average probability of observing the
``no error'' result $\bf 0$ in all $M$ cycles is approximately
$\exp(-tN/2T_1)$.

\section{Two-qubit error detection and correction for phase qubits
\label{simple qed section}}

In this section we propose and analyze the operation of two-qubit
error detection/correction protocols designed for experimental
implementation with the current technology of superconducting phase
qubits. We will discuss several similar protocols (including the QED
protocol for energy relaxation); for all of them the goal is to
preserve an arbitrary initial state $|\psi_{\rm
in}\rangle=\alpha|0\rangle+\beta|1\rangle$ of a qubit.

    The first procedure (which we will mostly consider) is
shown in Fig.\ \ref{two qubit qed figure}; it is designed to
preserve the state $|\psi_{\rm in}\rangle$ of the upper (main)
qubit. Encoding is performed by preparing the lower (ancilla) qubit
in the state $(|0\rangle +|1\rangle )/\sqrt{2}$ by starting with the
ground state $|0\rangle$ and using the $Y$-rotation over the angle
$\pi/2$ (denoted as $Y/2$), and then applying the controlled-$Z$
(C$Z$) gate between the two qubits. [Note that the C$Z$ gate is the
natural entangling operation for the phase qubits
\cite{Martinis-09,Matteo-11,Yamamoto-10}.] This produces the
entangled two-qubit wavefunction
    \be
[\alpha|0\rangle\otimes(|0\rangle +
|1\rangle)+\beta|1\rangle\otimes(|0\rangle-|1\rangle)]/\sqrt{2},
    \label{exp-enc}\ee
where the leftmost entry represents the main qubit. After encoding,
the decoherence process is simulated by applying a unitary rotation
to one of the qubits. For this encoding we consider a set of four
possible rotations: $R_{1}^{X}(2\theta)$, $R_{1}^{Y}(2\theta)$,
$R_{2}^{Y}(2\theta)$, and $R_{2}^{Z}(2\theta)$, where the subscript
indicates the qubit number (1 for the main qubit), the superscript
is the rotation axis on the Bloch sphere, and the argument $2\theta$
is the rotation angle on the Bloch sphere (the corresponding
rotation angle in the wavefunction language is $\theta$).

\begin{figure}[ptb]
\centering
\includegraphics[width=3.0in]{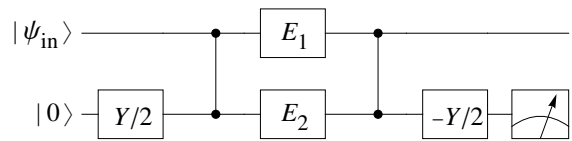} \caption{Two-qubit
experimental protocol for realizing quantum error
detection/correction. Notations $E_{1}$ and $E_{2}$ represent the
four detectable error rotations: $R_{1}^{X}(2\theta)$,
$R_{1}^{Y}(2\theta)$, $R_{2}^{Y}(2\theta)$, and
$R_{2}^{Z}(2\theta)$. Notations $Y/2$ and $-Y/2$ represent
$R^{Y}(\pi/2)$ and $R^{Y}(-\pi/2),$ respectively.} \label{two qubit
qed figure}
\end{figure}

After the error rotation has been applied, the resultant state is
decoded by inverting the encoding operation, and the ancilla qubit
is measured in the computational basis. In the absence of error
rotation, the state after decoding is
$(\alpha|0\rangle+\beta|1\rangle) \otimes |0\rangle $, so that the
initial state of the main qubit is restored and the measurement
result for the ancilla qubit is $0$. The error rotation disturbs the
final state, which probabilistically changes the measurement result
to $1$ and also changes the final state of the main qubit.

\subsection{Analysis of the ideal case}

Let us start with analyzing the effect of the error rotation
$R_{1}^{X}(2\theta)$ ($X$-rotation of the main qubit). It transforms
the encoded state (\ref{exp-enc}) into the state
    \begin{eqnarray}
&& [\alpha(\cos\theta|0\rangle-i\sin\theta|1\rangle) \otimes
(|0\rangle+|1\rangle)
    \nonumber \\
 && + \beta(-i\sin\theta|0\rangle +
\cos\theta|1\rangle)  \otimes (|0\rangle-|1\rangle)]/\sqrt{2},
    \end{eqnarray}
which after decoding (but before measurement) becomes
    \begin{equation}
\cos\theta\, |\psi_{\rm in}\rangle|0\rangle + i\sin\theta\,
X|\psi_{\rm in}\rangle|1\rangle ,
    \label{exp-final-1}\end{equation}
    where $X$ is the Pauli-$X$-matrix transformation
\cite{N-C}. It is clear that we obtain ancilla measurement result
$0$ with probability $\cos^{2}\theta$, and then the state of the
main qubit is restored to $|\psi_{\rm in}\rangle$, or obtain result
$1$ with probability $\sin^{2}\theta$, which leaves the main qubit
in the state $X|\psi_{\rm in}\rangle.$ [In this section we use the
standard quantum computing notations \cite{N-C}, in which the Pauli
matrices act on column vectors with the upper element corresponding
to the state $|0\rangle$. Note that for one-qubit wavefunctions
$R^X(\pi)=-iX$.]

    In quantum error detection we select only result 0, and this
gives the perfect state preservation fidelity, $F^{\rm qed}_{\rm
st}=1$, for any initial state. We can also use the approach of
quantum error correction and apply the $X$ gate [i.e. $R^X(\pi)$] to
the main qubit when the error result 1 is measured. This produces
the initial state $|\psi_{\rm in}\rangle$ for both measurement
results with perfect fidelity, $F^{\rm qec}_{\rm st}=1$. Therefore,
the QED and QEC fidelities averaged over the Bloch sphere are also
perfect,
    \be
F^{\rm qed}_{\rm av}=F^{\rm qec}_{\rm av}=1.
    \label{exp-qed-qec-perfect}\ee
Notice, however, that for QEC we had to know that an error is due to
the $X$-rotation applied to the first qubit. This is different from
``real'' error correction, in which we do not know the type of
error, but is acceptable for a demonstration experiment.

    Let us also calculate the storage fidelity if the measurement
result is ignored (or, equivalently, the ancilla qubit is not
measured). From Eq.\ (\ref{exp-final-1}) we obtain the state
fidelity for the main qubit $F^{\rm ign}_{\rm st}=\cos^{2}\theta +
\sin^{2}\theta \, \langle\psi_{\rm in}|X|\psi_{\rm in}\rangle^{2}$,
which after averaging over the Bloch sphere becomes
    \be
F^{\rm ign}_{\rm av}=\cos^{2}\theta + (\sin^{2}\theta)/3.
    \label{exp-ign}\ee
Note that if the rotation $R^{X}(2\theta)$ is applied to a qubit
without encoding, then the average fidelity is still given by Eq.\
(\ref{exp-ign}), so the encoding with ignored measurement result (or
no measurement) does not affect the average preservation fidelity
(moreover, it does not affect the state fidelity for any initial
state).

    Now let us analyze in a similar way the case when the error is
introduced by the $Y$-rotation of the main qubit,
$R_{1}^{Y}(2\theta)$. Then the two-qubit state before the
measurement is
    \be
\cos\theta\, |\psi_{\rm in}\rangle|0\rangle + i\sin\theta\,
Y|\psi_{\rm in}\rangle|1\rangle ,
    \ee
so that the measurement result 0 still restores the initial state
$|\psi_{\rm in}\rangle$ of the main qubit, while for the measurement
result 1 the state of the main qubit is $Y|\psi_{\rm in}\rangle$,
thus requiring the $Y$-gate correction [i.e.\ $R^Y(\pi)=-iY$]. The
QED and QEC fidelities are still perfect, Eq.\
(\ref{exp-qed-qec-perfect}), while the fidelity with ignored result
is still given by Eq.\ (\ref{exp-ign}). Note that the correcting
$Y$-gate is different from the correcting $X$-gate in the previous
case, so we need to know the type of the error to apply the proper
correction (in a demonstration experiment the error rotation is
applied intentionally, so its type is obviously known).

    Now let us consider the error due to the $Y$-rotation of the
ancilla qubit, $R_{2}^{Y}(2\theta)$. Then the state before the
measurement is
\begin{equation}
\cos\theta|\psi_{\rm in}\rangle|0\rangle + \sin\theta Z|\psi_{\rm
in}\rangle|1\rangle ,
\end{equation}
and therefore in the case of measurement result 1 the $Z$-gate
correction is needed to restore $|\psi_{\rm in}\rangle$, while for
the measurement result 0 no correction is needed. Equations
(\ref{exp-qed-qec-perfect}) and (\ref{exp-ign}) are still valid.

    Finally, for the $Z$-rotation of the ancilla qubit,
$R_{2}^{Z}(2\theta)$, the state before the measurement is
    \begin{equation}
|\psi_{\rm in}\rangle (\cos\theta\, |0\rangle +i\sin\theta\,
|1\rangle ) .
    \end{equation}
The final state of the main qubit is insensitive to this rotation,
and therefore no correction is needed for both measurement results.
In this case Eq.\ (\ref{exp-qed-qec-perfect}) is still valid, while
Eq.\ (\ref{exp-ign}) is replaced by $F^{\rm ign}_{\rm av}=1$.

We have discussed the effect of four error rotations:
$R_{1}^{X}(2\theta)$, $R_{1}^{Y}(2\theta)$, $R_{2}^{Y}(2\theta)$,
and $R_{2}^{Z}(2\theta)$.
    The two remaining rotations, $R_{1}^{Z}(2\theta)$ and
$R_{2}^{X}(2\theta)$, gradually change the final state of the main
qubit (both produce its $Z$-rotations) but always produce final
state $|0\rangle$ of the ancilla qubit. Therefore, these errors are
undetectable and are excluded from our set of error rotations.

    As discussed above, for QEC we need to know which one out
of four error types has been applied. In contrast, for QED we do not
need to know the error type; for all of them the measurement result
0 indicates the perfect state of the main qubit. Moreover, for QED
these types of error rotations can be applied simultaneously, as
long as the rotation angles are relatively small, to make negligible
the second-order terms (which in the QEC/QED language correspond to
double-errors).

    It is most natural to view the analyzed procedure as a QED
protocol. However, we would like to emphasize that its
interpretation as a QEC protocol is also possible: the proper
correction is possible when we know which error process is applied.
(In the existing QEC experiments the types of allowed errors are
almost always limited; in our protocol the number of allowed types
is further reduced to one out of four.) Most importantly, our simple
two-qubit protocol demonstrates the main ``miracle'' of QEC, that
continuous quantum errors can be transformed into discrete errors,
and then corrected.

\subsection{Realization using phase qubits}

    So far we considered the ideal case when there is no physical
decoherence of qubits, and the loss of fidelity is only due to
intentional rotations of the qubit states. In this subsection we
discuss a more realistic experimental situation, with added
decoherence during the protocol. We will have in mind the
present-day technology of superconducting phase qubits
\cite{Martinis-09,Matteo-11,Yamamoto-10}.

    Note that the phase qubit technology provides a high-fidelity
measurement (about 95\% \cite{Martinis-09}, so we consider it
perfect in the simulations); however, it takes a significant time to
read out the measurement result (longer than the qubit decoherence
time). While this is not a problem for the QED, the QEC at present
cannot be done in real time. Nevertheless, there is a simple way to
go around this difficulty in an experiment. The resulting state of
the main qubit is measured by using the quantum state tomography
(QST), so the experiment is necessarily repeated many times. It is
easy to separate the QST data for ancilla measurement results 0 and
1. In this way two different density matrices of the main qubit are
obtained for ancilla measurement 0 and 1. For the result 1 it is
then easy to calculate the density matrix after the correcting
operation (if it were applied in real time). Finally adding the two
density matrices (with weights equal to the probabilities of ancilla
measurement results), the qubit density matrix for the QEC procedure
is obtained.

    In phase qubits \cite{Martinis-09,Matteo-11,Yamamoto-10}
the main sources of decoherence are single-qubit energy relaxation
(with $T_1$ on the order of 0.5 $\mu$s) and pure dephasing (with a
comparable or a little shorter dephasing time $T_{\varphi}$). The
decoherence is somewhat reduced in the RezQu architecture
\cite{Matteo-11,RezQu}, in which the quantum information is often
transferred between the phase qubits and resonators (resonators have
much longer $T_1$ and practically no pure dephasing).
    We have simulated the procedure of Fig.\ 4 in a simplified way,
which does not explicitly reproduce the RezQu implementation of the
protocol, but still uses a reasonable account of realistic
decoherence.

For simplicity for each qubit we assume $T_1=T_2$ (so that the pure
dephasing time is $T_\varphi = 2T_1$). We assume that single-qubit
rotations [including $R^{Y}(\pm\pi/2)$ of ancilla qubit, preparation
of the main qubit, and error rotations] take 10 ns each, C$Z$ gates
take 40 ns each, and there are 5 ns spacings between the operations.
Then the whole protocol of Fig.\ 4 (ending before measurement of
ancilla qubit and tomography of the main qubit) takes 135 ns. We
calculate the evolution of the two-qubit density matrix by breaking
the procedure into small time steps and applying energy relaxation
and pure dephasing to each qubit (for simplicity the C$Z$ gate is
simulated as a gradual accumulation of the phase, as would be for
the dispersive gate). We start with 6 initial states of the main
qubit [$|0\rangle$, $|1\rangle$, $(|0\rangle \pm
|1\rangle)/\sqrt{2}$, $(|0\rangle \pm i |1\rangle)/\sqrt{2}$,
labeled by index $j=1,\dots 6$ below], and from the final two-qubit
density matrices we calculate reduced unnormalized one-qubit density
matrices $\rho_{0,j}$ and $\rho_{1,j}$, corresponding to ancilla
measurement results 0 and 1 (the probabilities of these results are
then ${\rm Tr}\rho_{0,j}$ and ${\rm Tr}\rho_{1,j}$). The averaged
preservation fidelity with ignored measurement results is then (see
Appendix) $F_{\rm av}^{\rm ign}=(1/6)\sum_j {\rm Tr} [
(\rho_{0,j}+\rho_{1,j})\rho^{\rm in}_j]$, where $\rho^{\rm
in}_j=|\psi_j\rangle \langle \psi_j|$ is the unchanged initial
state. The averaged (weighted) QED fidelity is then $\widetilde
F_{\rm av}^{\rm qed}=\sum_j {\rm Tr} (\rho_{0,j} \rho^{\rm
ideal}_j)/\sum_j {\rm Tr} \rho_{0,j} $, and the QEC fidelity is
$F_{\rm av}^{\rm qec}=(1/6)\sum_j {\rm Tr} [
(\rho_{0,j}+\rho_{1,j}^{\rm corr})\rho^{\rm in}_j]$, where the
corrected density matrix $\rho_{1,j}^{\rm corr}$ is obtained from
$\rho_{1,j}$ by applying the ideal correcting operations ($X,Y,Z,I$)
discussed in the previous subsection.

    Figure 5 shows the average fidelities $\widetilde F_{\rm av}^{\rm qed}$
(solid lines), $F_{\rm av}^{\rm qec}$ (dotted lines), and $F_{\rm
av}^{\rm ign}$ (dashed lines), as functions of the rotation angle
$2\theta$ (in units of $\pi$) for the intentional $X$-rotation of
the main qubit, $R_{1}^X(2\theta)$. The three sets of lines are for
three values of $T_1=T_2$: 300 ns, 500 ns, and 700 ns. Note that we
present the average fidelities $F_{\rm av}$, but they can be easily
converted into the process matrix fidelities $F_\chi$ via Eq.\
(\ref{F-chi-av}). Also note that the range from 1/3 to 1 for $F_{\rm
av}$ (used for the vertical axis in Fig.\ 5) corresponds to the
range from 0 to 1 for $F_\chi$.

\begin{figure}[ptb]
\centering
\includegraphics[width=3.2in]{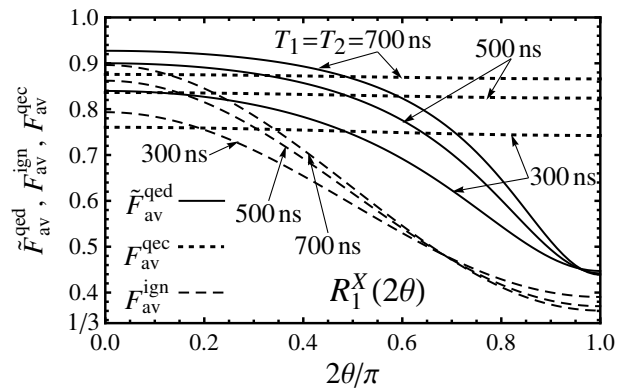}
\caption{Numerical results for the average QED fidelity $\widetilde
F_{\rm av}^{\rm qed}$ (solid lines), the QEC fidelity $F_{\rm
av}^{\rm qec}$ (dottes lines), and the fidelity $F_{\rm av}^{\rm
ign}$ with ignored ancilla measurement results (dashed lines), as
functions of the angle $2\theta$ of intentional $X$-rotation of the
main qubit, $R_{1}^{X}(2\theta)$. The simulated protocol of Fig.\ 4
has duration of 135 ns. We assume the qubits with $T_1=T_2=300$ ns,
500 ns, and 700 ns.
    } \label{qed many T1}
\end{figure}

\begin{figure}[ptb]%
\centering
\includegraphics[width=3.2in]{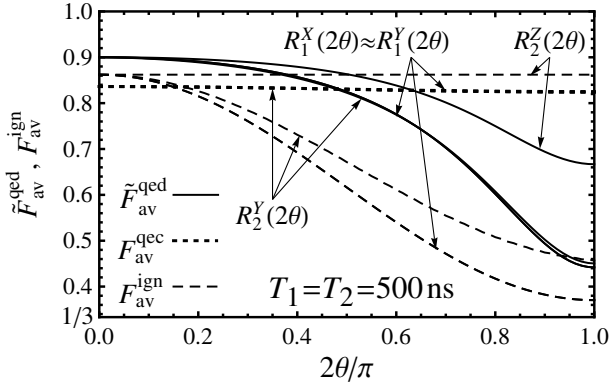}
   \caption{Same as in Fig.\ 5, but
for four types of intentional qubit state rotations:
$R_{1}^{X}(2\theta)$, $R_{1}^{Y}(2\theta)$, $R_{2}^{Y}(2\theta)$,
and $R_{2}^{Z}(2\theta)$. The qubits with $T_1=T_2=500$ ns are
assumed. Results for rotations $R_{1}^{X}(2\theta)$ and
$R_{1}^{Y}(2\theta)$ practically coincide.
}%
\label{qed four errors}%
\end{figure}

    From Fig.\ 5 we see that even for $T_1=T_2=300$ ns the QED
fidelity is significantly higher than the fidelity with ignored
measurement result (recall that the procedure duration is 135 ns).
The difference between $\widetilde F^{\rm qed}_{\rm av}$ and $F^{\rm
ign}_{\rm av}$ becomes larger for longer decoherence time (500 ns
and 700 ns). The QEC fidelity is below the QED fidelity (and even
below $F_{\rm av}^{\rm ign}$) for small $\theta$, but becomes above
$\widetilde F_{\rm av}^{\rm qed}$ and $F_{\rm av}^{\rm ign}$ at
large $\theta$.

    It is interesting to notice that $F_{\rm av}^{\rm ign}$ at
$2\theta \approx \pi$ is much closer to the ideal value 1/3 [see
Eq.\ (\ref{exp-ign})] than to the ideal value 1 at $2\theta \approx
0$. This property can be understood using the equivalent language of
the process fidelity $F_\chi={\rm Tr} ( {\chi}_{\rm desired}\chi)$
-- see Eq.\ (\ref{F-chi-av}). Since the desired operation is the
absence of evolution, $F_\chi=\chi_{II}$ in the standard notations
for the one-qubit 4$\times$4 matrix $\chi$
\cite{N-C,Blatt-11,Yamamoto-10,Matteo-11}; note that
$\chi_{II}+\chi_{XX}+\chi_{YY}+\chi_{ZZ}=1$. Ideally $\chi_{II}=1$
for $2\theta=0$ and $\chi_{XX}=1$ for $2\theta=\pi$. Since
decoherence spreads these ideal unity values to the three other
diagonal elements of $\chi$, we would expect that $F_\chi^{\rm ign}$
at $2\theta=\pi$ should be (very crudely) three times less than
$1-F_\chi^{\rm ign}$ at $2\theta=0$. This roughly corresponds to
what we see in Fig.\ 5.

    Figure 5 shows the results only for the $X$-rotation of the main
qubit, $R_{1}^X(2\theta)$. The results for all four considered
rotations, $R_{1}^X(2\theta)$, $R_{1}^Y(2\theta)$,
$R_{2}^Y(2\theta)$ and $R_{2}^Z(2\theta)$, are shown in Fig.\ 6 for
$T_1=T_2=500$ ns. The results for $X$ and $Y$-rotation of the main
qubit are practically indistinguishable from each other.
  The QED and QEC fidelities for
$Y$-rotation of the ancilla qubit are very close to the
corresponding fidelities for the rotation of the main qubit. For
$Z$-rotation of the ancilla qubit the operation with ignored
measurement coincides with the QEC operation (because no correction
is applied for measurement result 1), and the QED fidelity
$\widetilde F_{\rm av}^{\rm qed}$ is higher than $F_{\rm av}^{\rm
ign}=F_{\rm av}^{\rm qec}$ only at $2\theta \alt \pi/2$, and only by
a small amount. Obviously, the rotation $R_{2}^Z(2\theta)$ is not
good for demonstrating an advantage of this encoding, in contrast to
other rotations.

    Overall, from Figs.\ 5 and 6 we see that the current technology
of phase qubits is good enough for demonstrating the operation of
the considered two-qubit QED/QEC protocol. In an experiment, the
larger value of the QED fidelity in comparison with the case of
ignored measurement result is the demonstration that the QED
procedure is beneficial. Similarly, the QEC operation can also be
demonstrated (though with the caveat discussed in the previous
subsection).

\subsection{Related protocols}

\begin{figure}[ptb]
\centering
\includegraphics[width=3.2in]{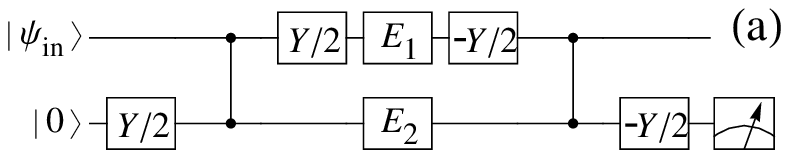}
\includegraphics[width=3.2in]{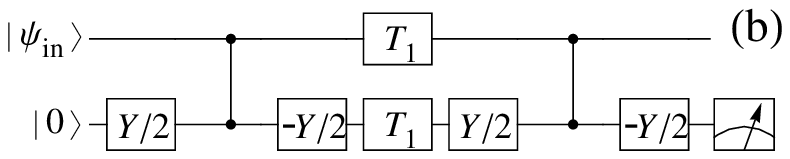}
    \caption{Modified
two-qubit QED/QEC algorithms. The protocol shown in (a)
detects/corrects errors due to rotations $R_{1}^{Y}$, $R_{1}^{Z}$,
$R_{2}^{Y}$, and $R_{2}^{Z}$; it can be used to protect from natural
pure dephasing of the qubits. The protocol in (b) is designed for
error rotations $R_{1}^{X}$, $R_{1}^{Y}$, $R_{2}^{X}$, and
$R_{2}^{Y}$. Therefore, it can be used as a QED procedure for errors
due to energy relaxation of both qubits (stored in resonators).}
    \label{modif}
\end{figure}

    The protocol of Fig.\ 4 can be easily modified to change the set
of four detectable/correctable error operations. For example, if we
desire protection from $Y$ and $Z$ rotations of both qubits [i.e.\
$R_{1}^{Y}(2\theta)$, $R_{1}^{Z}(2\theta)$, $R_{2}^{Y}(2\theta)$,
and $R_{2}^{Z}(2\theta)$], we can add $\pm \pi/2$ $Y$-rotations of
the main qubit before and after the error rotations -- see Fig.\
\ref{modif}(a). Such encoding also protects from natural pure
dephasing of both qubits.

    For protection from $X$ and $Y$ rotations of both qubits [i.e.\
$R_{1}^{X}(2\theta)$, $R_{1}^{Y}(2\theta)$, $R_{2}^{X}(2\theta)$,
and $R_{2}^{Y}(2\theta)$], we can add $\mp \pi/2$ $Y$-rotations of
the ancilla qubit before and after the errors -- see Fig.\
\ref{modif}(b). Such encoding can be used in the QED mode for the
energy relaxation of both qubits. (This procedure essentially
realizes the idea of Fig.\ 1 for two qubits; the only difference is
the encoding $\alpha |00\rangle -\beta |11\rangle$ instead of
encoding  $\alpha |00\rangle +\beta |11\rangle$ considered in Sec.\
II B.)

    In the RezQu architecture based on phase qubits \cite{Matteo-11,RezQu},
the protocol of Fig.\ \ref{modif}(b) can be efficiently used for
storing the information in the resonators. We have simulated the
operation of this protocol, assuming that the encoding/decoding is
done with the phase qubits having relaxation times $T_1=T_2$, while
in between encoding and decoding the information is moved to
resonators for a relatively long storage. The procedure (without the
storage time) is slightly longer than what was considered in the
previous subsection: 155 ns instead of 135 ns (we still do not
simulate explicitly the move operations between the qubits and
resonators).
    Solid lines in Fig.\ 8 show the corresponding QED fidelities
$\widetilde F_{\rm av}^{\rm qed}$ as functions of the single-qubit
energy relaxation probability $p=1-\exp(-t_{\rm storage}/T_1^{\rm
resonator})$ during the storage (in experiment \cite{Matteo-11}
$T_1^{\rm resonator} \gg T_1$, though our results do not need this
assumption). The QEC operation is impossible in this case (for real
energy relaxation in resonators); however, as seen from Fig.\ 8, the
QED operation can be reliably demonstrated experimentally.

\begin{figure}[ptb]
\centering
\includegraphics[width=3.1in]{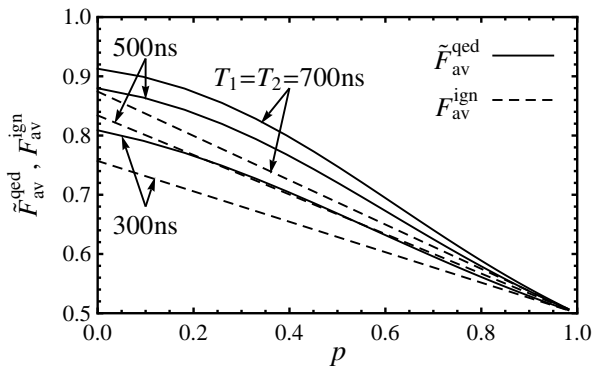}
 \caption{Average QED fidelity $\widetilde F_{\rm av}^{\rm qed}$
(solid lines) for the two-qubit protocol of Fig.\ \ref{modif}(b), as
a function of the single-qubit energy relaxation probability
$p=1-\exp(-t_{\rm storage}/T_1^{\rm resonator})$ during information
storage in resonators. Dashed lines show average fidelity $F_{\rm
av}^{\rm ign}$ when the measurement result is ignored. The
encoding/decoding is done with phase qubits having significantly
shorter relaxation times $T_1=T_2$ (300 ns, 500 ns, and 700 ns); the
assumed duration of the procedure (excluding storage time) is 155
ns.
    }
    \label{exp-T1}
\end{figure}

\section{Conclusion}

    In Sec.\ II we have analyzed the performance of $N$-qubit repetitive
quantum codes in the presence of energy relaxation. As expected,
these codes are not usable for quantum error correction. However,
they can be used for quantum error detection. The best QED
performance for weak energy relaxation is provided by the 3-qubit
repetitive encoding [see Eq.\ (\ref{Nq-qed-approx}) and Fig.\ 3(a)],
while the 2-qubit encoding is sufficient and gives only slightly
lower fidelity [see Eq.\ (\ref{2q-av-qed-approx})]. We have found
that the main contribution to the QED infidelity for $N\geq 3$ comes
from the non-unitary change of the quantum state in the case when no
relaxation happens. Therefore, the QED infidelity can be strongly
decreased if the QED algorithm is complemented with partial quantum
measurement or, alternatively, if the protocol is divided into the
even number of cycles and complemented with $\pi$-pulses in between
(this resembles dynamical decoupling, though only superficially). In
this case the fidelity improves with dividing the total duration
into a larger number of cycles and using more qubits for the
encoding [see Eq.\ (\ref{Nq-cycles})].

    Note that the QED fidelity cannot be introduced in the usual way
[as $F_\chi={\rm Tr} (\chi_{\rm desired}\chi )$] via the process
$\chi$-matrix. In the analysis we have used the state fidelity,
averaged over the Bloch sphere, with the usual conversion into
$F_\chi$ via Eq.\ (\ref{F-chi-av}). For the state fidelity averaged
with the weight proportional to the selection probability (denoted
$\widetilde F_{\rm av}^{\rm qed}$) the usual trick of averaging over
the 6 initial states can be used (see Appendix), while for the state
fidelity averaged with uniform weight (denoted $F_{\rm av}^{\rm
qed}$) we had to average over the Bloch sphere explicitly, using
Eqs.\ (\ref{Bloch-av-1})--(\ref{Bloch-av-8}). In the analysis we
used unraveling of the decoherence evolution into the ``error
scenarios'', which is in general similar to the standard approach
used in the quantum error correction, but is different in the way
that each scenario describes a non-unitary process.

    In Sec.\ III we have considered simple two-qubit protocols of
quantum error detecton/correction, suitable for present-day
experiments with superconducting phase qubits \cite{Matteo-11}. In
the protocol of Fig.\ 4 the errors are simulated by intentional
unitary rotations of the qubit states (two types of rotations for
each qubit). In this case not only the QED, but also the QEC
operation is possible if we know the applied type of error rotation.
Most importantly, this experiment would demonstrate the QEC
``miracle'' of converting continuous quantum errors into discrete
errors, which are then correctable.
 The numerical simulations (Figs.\ 5 and 6) with
account of decoherence during the protocol show that the
experimental QED and QEC fidelities are expected to be significantly
higher than the fidelity with ignored result of the ancilla qubit
measurement. Therefore, the QED and QEC benefits can be demonstrated
experimentally.

    A slightly different protocol, shown in Fig.\ 7(b), can be used
as a QED procedure for errors due to natural energy relaxation of
qubits stored in resonators of a RezQu-architecture device
\cite{Matteo-11,RezQu} based on phase qubits. The numerical
simulations (Fig.\ 8) show that such experiment can also be realized
with the present-day technology, demonstrating the benefits of
encoding a logical qubit in several (two in this case) physical
qubits. While the measurement-free QEC experiment has been recently
realized with superconducting transmon qubits \cite{Reed-12}, the
experiments proposed and analyzed in this paper would be the first
measurement-based QED/QEC protocols realized with superconducting
qubits.

\vspace{0.3 cm}

    The authors thank John Martinis, Andrew Cleland, and Matteo
Mariantoni for fruitful discussions.  The work was supported by the
IARPA under ARO Grant No.\ W911NF-10-1-0334.

\appendix
\section{}

    In this appendix we prove that no additional unitary operation can
improve the fidelity of the protocol discussed in Sec.\ II (for
2-qubit or $N$-qubit encoding) in the case of ``no error''
measurement result (0 or $\bf 0$). We also prove that for a
measurement result which indicates an error, the optimal correction
is either identity or the $\pi$-pulse, exchanging $|0\rangle$ and
$|1\rangle$. Along the way we also discuss the trick
\cite{Nielsen-02,Bowdrey-02} of using only 6 initial states for
averaging the state fidelity over the Bloch sphere.

    Let us first consider an arbitrary (not necessarily
trace-preserving) linear one-qubit quantum operation, which
transforms initial states $|1\rangle$, $|0\rangle$, $|\pm\rangle
\equiv(|0\rangle \pm |1\rangle)/\sqrt{2}$, $|\pm i\rangle \equiv
(|0\rangle \pm i|1\rangle)/\sqrt{2}$ into the density matrices
$\rho_0$, $\rho_1$, $\rho_\pm$, $\rho_{\pm i}$. The center of the
Bloch sphere $(|0\rangle \langle 0| + |1\rangle \langle 1|)/2=I/2$
is transformed into $\rho_c$. Because of the linearity, only four
linearly independent initial states are sufficient to define the
operation. So, for an initial state with the Bloch sphere
coordinates $\{x,y,z\}$,
    \be
    \rho_{\rm in}= ( I + x X+yY +zZ)/2,
    \ee
where $\{X,Y,Z\}$ are the Pauli matrices ($x=1$ corresponds to
$|+\rangle$, $y=1$ corresponds to $|+i\rangle$, $z=1$ corresponds to
$|0\rangle$), the final state is
   \be
    \rho_{\rm fin}= \rho_c + x (\rho_+-\rho_c) + y (\rho_{+i}-\rho_c)
    + z (\rho_0-\rho_c) .
    \ee

    To compare this operation with a unitary $U$, we calculate the
state fidelity ${\rm Tr} (\rho_{\rm fin} \rho_{\rm fin}^U)$ (the
superscript $U$ in a notation means that it relates to the unitary
$U$),
    \begin{eqnarray}
F_{\rm st} = {\rm Tr} \{ [\rho_c + x (\rho_+-\rho_c) + y
(\rho_{+i}-\rho_c)
    + z (\rho_0-\rho_c)]
    \nonumber \\
 \times[\rho_c^U + x (\rho_+^U -\rho_c^U) + y (\rho_{+i}^U-\rho_c^U)
    + z (\rho_0^U-\rho_c^U)] \}. \quad
    \label{F-st-app}\end{eqnarray}
Note that $\rho_c^U=I/2$, since a unitary operation does not change
the Bloch sphere center. In averaging $F_{\rm st}$ over the Bloch
sphere we average over the coordinates $\{x,y,z\}$ and use the
obvious relations
$\overline{x}=\overline{y}=\overline{z}=\overline{xy}=\overline{xz}=
\overline{yz}=0$,
$\overline{x^2}=\overline{y^2}=\overline{z^2}=1/3$, this obtaining
   \begin{eqnarray}
&& \hspace{-0.5cm}  \overline{F} = \frac{1}{2} {\rm Tr} \rho_c
+\frac{1}{3} {\rm Tr} [(\rho_+-\rho_c)(\rho_+^U -\rho_c^U)
    \nonumber \\
&& +(\rho_{+i} -\rho_c)(\rho_{+i}^U -\rho_c^U)+(\rho_0
-\rho_c)(\rho_0^U -\rho_c^U) ].  \quad
    \label{F-av-app}\end{eqnarray}
Note that in general we deal here with non-normalized density
matrices, in contrast to the formalism used in Sec.\ II. Therefore,
compared with notations of Sec.\ II, $\overline F = F_{\rm av}=
\widetilde F_{\rm av}$ only for a trace-preserving operation, while
for a non-trace-preserving operation $\overline{F}=\widetilde F_{\rm
av}\overline{P}$, where $\overline{P}$ is the average probability of
selection [see Eq.\ (\ref{F-av-tilde-F})], and there is no direct
relation between $\overline{F}$ and $F_{\rm av}$.

    Using Eq.\ (\ref{F-av-app}) it is easy to see why averaging over the
Bloch sphere is equivalent to averaging over only 6 initial states:
$|0\rangle$, $|1\rangle$, $|\pm\rangle$, and $|\pm i\rangle$. The
state fidelity $F_+$ for the initial state $|+\rangle$ is given by
Eq.\ (\ref{F-st-app}) with $x=1$ and $y=z=0$. The state fidelity
$F_-$ for the initial state $|-\rangle$ is similar, but $x=-1$. It
is easy to obtain the sum, $F_+ +F_- = {\rm Tr} \rho_c + 2 {\rm Tr}
[ (\rho_+ - \rho_c)(\rho_+^U-\rho_+^U)]$, which is similar to the
terms in Eq.\ (\ref{F-av-app}). Similarly finding the sums $F_{+i}
+F_{-i}$ and $F_0 +F_1$, we obtain \cite{Nielsen-02,Bowdrey-02}
    \be
    \overline{F} = (F_0+F_1+F_++F_-+F_{+i}+F_{-i})/6.
    \ee
Note that this relation remains valid for non-trace-preserving
operations, when we are working with a linear operation and
non-normalized states. The same six-point-averaging relation is
valid for the average probability of selection $\overline P$,
because $P={\rm Tr} \rho_{\rm fin}$ and therefore $\overline P={\rm
Tr} \rho_c$ (even two-point averaging is sufficient for $\overline
P$, when we choose two opposite points on the Bloch sphere).
Therefore the six-point-averaging trick is useful for finding
$\widetilde F_{\rm av}=\overline{F}/\overline{P}$.

    Now let us discuss why an additional unitary cannot improve the
QEC protocols of Sec.\ II when the ``no error'' measurement result 0
(or {\bf 0}) is obtained. The final state in this case is an
incoherent mixture of the results of two linear operations:
    \be
    \alpha |0\rangle +\beta |1\rangle \rightarrow \alpha |0\rangle +
    k \beta |1\rangle , \,\,\,\,
       \alpha |0\rangle +\beta |1\rangle \rightarrow \tilde{k} \beta
       |0\rangle ,
    \label{proc-app}\ee
where the real positive constants $k$ and $\tilde k$ should
obviously satisfy inequality $k^2+\tilde k^2 \leq 1$. For this
operation it is easy to find explicitly
    \begin{eqnarray}
   \rho_c =(1+\tilde k^2+k^2)I/4 + (1+\tilde k^2-k^2)Z/4 ,
    \qquad \qquad   \\
    \rho_+ = \rho_c + kX/2 , \,\, \rho_{+i}=\rho_c + kY/2, \,\,
\rho_0=(Z+I)/2. \quad
    \end{eqnarray}
Then using Eq.\ (\ref{F-av-app}) we obtain
   \begin{eqnarray}
&& \hspace{-0.5cm}  \overline{F} = \frac{1}{4} (1+k^2+\tilde k^2) +
\frac{1}{6} {\rm Tr} [k X (\rho_+^U -\rho_c^U)
    \nonumber \\
&& +k Y(\rho_{+i}^U -\rho_c^U) +\frac{1-\tilde k^2+k^2}{2} Z
(\rho_0^U -\rho_c^U) ].
    \end{eqnarray}
(Note that comparing the operation with $U$ we assume the correction
operation $U^\dagger$.) Optimizing each term under the trace over
the unitary $U$ separately, we see that the first term is maximized
by unitaries, which transform $|+\rangle \rightarrow |+\rangle$; the
maximum for the second term is achieved when $|+i\rangle \rightarrow
|+i\rangle$, and the maximum for the third term is achieved when
$|0\rangle \rightarrow |0\rangle$ (note that $k\geq 0$ and $1-\tilde
k^2+k^2 \geq 0$). Since the no-evolution operation satisfies all
these conditions, it provides the maximum fidelity,
    \be
   U_{\rm best}=I, \,\,\, \overline F = \frac{1+k+k^2+\tilde k^2/2}{3}.
    \ee
Note that the average probability of the process (\ref{proc-app}) is
$\overline{P}=(1+k^2+\tilde k^2)/2$, so $\widetilde F_{\rm
av}=(2/3)(1+k+k^2+\tilde k^2)/(1+k^2+\tilde k^2)$. In particular,
this is an alternative way of deriving Eq.\ (\ref{2q-av-t2}) by
using $k=\sqrt{1-p_1}\sqrt{1-p_2}$ and $\tilde k=\sqrt{p_1p_2}$.

    Now let us discuss what is the optimal unitary correction
operation after obtaining the measurement result 1 in 2-qubit
encoding or any result except $\bf 0$ in $N$-qubit encoding. Then
the resulting state is an incoherent mixture of two linear
operations:
    \be
    \alpha |0\rangle +\beta |1\rangle \rightarrow k \beta |1\rangle ,
    \,\,\,\,
       \alpha |0\rangle +\beta |1\rangle \rightarrow \tilde{k} \beta
       |0\rangle .
    \label{proc2-app}\ee
Finding explicitly the resulting density matrices
    \be
    \rho_0=0, \,\,\, \rho_{c}=\rho_+=\rho_{+i}= \frac{\tilde k^2}{2}|0\rangle
    \langle 0|+ \frac{k^2}{2}|1\rangle    \langle 1|,
    \ee
we obtain from Eq.\ (\ref{F-av-app})
    \be
    \overline F= \frac{k^2+\tilde k^2}{4} + \frac{k^2-\tilde k^2}{12} \,
     {\rm Tr} [Z(\rho_0^U-\rho_c^U)] .
    \ee
Therefore if $k\geq \tilde k$, then the maximum fidelity
$\overline{F}_{\rm max}=(2k^2+\tilde k^2)/6$ is achieved for any
unitary $U$, which does not change $|0\rangle$ (same for the
correcting operation $U^\dagger$). However, if $k\leq \tilde k$,
then the optimal $U$ transforms $|0\rangle \rightarrow |1\rangle$
(same for $U^\dagger$) and $\overline{F}_{\rm max}=(k^2+2\tilde
k^2)/6$.

\end{document}